\documentclass[
    ,final            
  ]
  {aipproc}

\layoutstyle{6x9}

\usepackage{color}

\newcommand{\e}{{\rm e}}
\newcommand{\rmd}{{\rm d}}
\newcommand{\rmi}{{\rm i}}

\newcommand{\up}{\uparrow}
\newcommand{\dn}{\downarrow}
\newcommand{\tr}{\hbox{tr}}

\newcommand{\half}{{\textstyle{\frac{1}{2}}}}

\newcommand{\de}{\delta}

\newcommand{\Om}{\Omega}

\begin{document}

\title[]
{Poor qubits make for rich physics: \\
{noise-induced quantum Zeno effects, \\ and noise-induced Berry phases.}}


\author{Robert S. Whitney}{
  address={Institut Laue-Langevin, 6 rue Jules Horowitz, B.P. 156,
         38042 Grenoble, France.}
}

\begin{abstract}
We briefly review three ways that environmental noise can slow-down
(or speed-up) quantum transitions; (i) Lamb shifts, (ii) over-damping and
(iii) orthogonality catastrophe.
We compare them with the quantum Zeno effect induced by 
observing the system. 
These effects are relevant to poor qubits (those strongly coupled to
noise).
We discuss Berry phases generated by the orthogonality catastrophe, 
and argue that noise may make it {\it easier} to observe Berry phases.
\end{abstract}

\maketitle


Our presentation for ICNF 2009
is based on our work entitled {\it ``Noise suppressing non-adiabaticity: observing a Berry phase alone''} \cite{whit-BP}. 
Here, we review the literature about the ways in which environmental noise can 
change the rate of quantum transitions (noise-induced quantum Zeno effects),
and then briefly discuss the applications of these ideas to noise-induced 
Berry phases, such as those in Ref.~\cite{whit-BP}.

Since qubits have been created and studied in experiments, 
theoretical descriptions of  dissipation in two-level system have a renewed
relevance.
A ``good'' qubit is a fully controllable two-level system which 
is sufficiently weakly coupled to environmental noise that it experiences
only weak dissipation (often well-described by
the Bloch-Redfield equation \cite{Bloch-Redfield57}).  
For quantum computing, qubits need to have
coherent oscillations with a quality factor of $10^4$ or more
(ten-thousand coherent oscillations before decaying).
Only then will errors be rare enough that they could be fixed by
error correction codes. 

In contrast a ``poor'' qubit is a controllable two-level system
which is strongly coupled to its environment, and thus experiences strong
dissipation.  The theory for this was discussed 
by Leggett {\it et al} \cite{Leggett}, but there are still many open
questions.  It is worth noting that even the best qubits
are  only  ``good'' for certain values of their parameters
(values where the noise couples quadratically rather than linearly to
superpositions of the two-levels). The qubits rapidly become ``poor'' 
away from these special points \cite{Saclay-qubit}.
Our aim here is to reveal a little of the rich physics
of such poor qubits.

Environmental noise (of classical or quantum origin) causes dissipation in quantum systems.
We  start this brief review with two very general and open questions.
\begin{itemize}
\item[I)]
What can a dissipative quantum system do that a non-dissipative system
cannot?
\item[II)]
Could noise be used to {\it control} a quantum system
in a manner that 
cannot be achieved using traditional Hamiltonian manipulation?
\end{itemize}
The first question is motivated by the observation that the density matrix
of a dissipative two-level system, $\rho$, 
is described by {\it three} independent parameters, while 
 a non-dissipative two-level system
is described by only {\it two}.
For example, one can write
\begin{eqnarray}
\rho = {1\over 2}
\left(\begin{array}{cc} 1+s_z & s_x -\rmi s_y \\ s_x +\rmi s_y & 1-s_z 
\end{array}\right)
\end{eqnarray}
where $s_x$, $s_y$ and $s_z$ are real numbers.
For non-dissipative (Hamiltonian) 
dynamics, all states must have a purity $P= \tr[\rho^2]=1$, which means that
$s_x^2+s_y^2+s_z^2=1$.
In contrast, for dissipative dynamics we only require that
$s_x^2+s_y^2+s_z^2 \leq 1$. This suggests that dissipation 
makes the dynamics much richer.
In this article we discuss some effects of dissipation, 
however it is not clear to us whether all possible effects are known.

Our second question, which is potentially relevant for quantum information processing, can definitely be answered by ``yes''!
Our presentation at this conference gives one example of this
(see the section below about Berry phases and Ref.~\cite{whit-BP}). 
A more obvious example is 
{\it thermalization}; which drives the system to a
single well-defined state, independent of the system's initial state.
For example, coupling the two-level system to a zero temperature environment
will cause it to decay to its ground state, irrespective of what 
state (mixed or pure) it was in before \cite{footnote:env-increase-P}.  
No non-dissipative evolution
(Hamiltonian evolution) can do this. 
We wonder if other examples exist?

\section{Quantum Zeno effects due to noise} 

In the quantum Zeno effect \cite{Zeno-effect}, 
the transitions of a quantum system are slowed down
(or stopped) by the fact one is observing the system.
It is the quantum mechanical equivalent of saying that 
``a watched pot never boils''
(i.e. a watched system never makes transitions).
However, there is an analogy between observing a system (projective 
measurements of its state) 
and environmental noise. 
Indeed one can model a measurement device as a large object with 
many degrees of freedom interacting with the system. 
Thus one can ask whether environmental-noise induces a quantum Zeno effect,
and (if so) what its nature is.
In fact, there are a number of noise-induced Zeno effects, 
i.e.~noise slows-down quantum transitions in a variety of
(qualitatively different) ways.
Also, certain noise can 
slightly speed up transitions (a weak anti-Zeno effect).

\vskip 3mm
{\bf Traditional quantum Zeno effect and anti-Zeno effect (due to observations) :}
Imagine a spin-half which is $\left|\up\right\rangle$ along the $z$-axis,
in an $x$-axis magnetic-field, so its Hamiltonian is  
\begin{eqnarray}
{\cal H}_{\rm spin} = -\half B_x 
\left(\begin{array}{cc} 0 & 1 \\ 1 & 0 \end{array}\right).
\end{eqnarray}
If the spin is left to evolve, it precesses around the $x$-axis, so that
after a time $\pi/B_x$ the spin will be $\left|\dn\right\rangle$.
However what if we observe the spin's z-axis polarization
at regular intervals $\tau_{\rm obs}$ much less than the time $B_x^{-1}$?
The observation is modelled by projecting out the spin-state
along the $z$-axis as follows:
\begin{eqnarray}
\left(\begin{array}{c} u \\ v \end{array}\right)
\longrightarrow 
\left\{ {\left(\begin{array}{c} 1 \\ 0 \end{array}\right) \hbox{ with prob.$=|u|^2$} 
\atop \left(\begin{array}{c} 0 \\ 1 \end{array}\right) \hbox{ with prob.$=|v|^2$} }
\right.
\end{eqnarray}
A simple calculation \cite{Zeno-effect} shows that for 
$B_x\tau_{\rm obs} \ll 1$,
the probability to find the spin up during all observations up to a time $t$
is $\exp[-B_x^2\tau_{\rm obs}t/4]$.
Thus rapid observations greatly reduce the 
spin's transition-rate.
The anti-Zeno effect
is the reverse; transitions are speeded up by 
the fact one is observing the system. The circumstance under which this occurs
is nicely explained in Ref.~\cite{anti-Zeno-effect}.

\begin{figure}
  \includegraphics[width=0.8\textwidth]{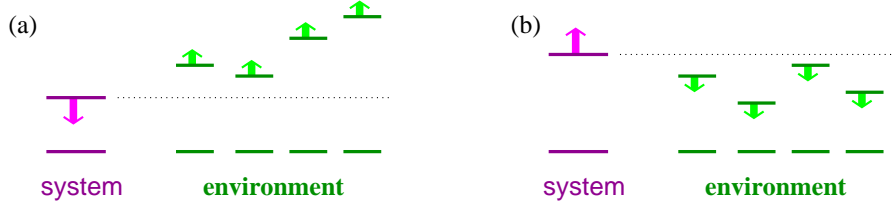}
  \caption{
The Lamb shift.
In (a) the system-levels have a smaller gap than the environment-levels.
As one turns-on a system-environment interaction, there is 
level-repulsion which {\it reduces} the system's gap (the purple arrow
on the system's upper state).  
In (b) the system-levels have a bigger gap than the environment-levels, 
so level-repulsion {\it increases} the system's gap.  
In most cases the environment has gaps both bigger and smaller than the 
system's 
gap, then the Lamb shift must be calculated.
\label{Fig:Lamb}}
\end{figure}

\vskip 3mm
{\bf Weak quantum Zeno and anti-Zeno effects due to Lamb shifts :}
In 1947 Lamb noted that atomic Hydrogen's levels are slightly shifted with 
respect to the predictions of Schrodinger (or Dirac) quantum mechanics.
The same year, Bethe explained this shift in terms of the interaction
between the electron and the photonic excitations of the vacuum.
Similar Lamb shifts occur whenever a system is coupled to an environment, 
and can be thought of in terms of level-repulsion, see Fig.~\ref{Fig:Lamb}. 
One can interprete this as a weak Zeno or anti-Zeno effect,
in which the spin's coherent oscillations are either slightly slowed-down, 
or slightly speeded-up by a quantum environment.

\vskip 3mm
{\bf Quantum Zeno effect due to over-damped spin dynamics}
There is a very pretty model of dissipative quantum mechanics in which one
averages the quantum dynamics over classical white-noise.
The fact the noise is white (meaning it contains all frequencies) is equivalent
to saying that it is $\de$-correlated in time.  This means one can time-slice
on a timescale much less than that associated with the system dynamics, 
and then average over the noise in each slice independently.
If the spin evolves under the Hamiltonian ${\cal H}_{\rm spin}$
above, and we turn on noise along the $z$-axis, 
one finds that \cite{Blanchard}
\begin{eqnarray}
{\rmd \over \rmd t} 
\left(\begin{array}{c} s_x \\ s_y \\ s_z \end{array}\right)
&=&
\left(\begin{array}{ccc}
-2\Gamma  & 0 & 0 \\
0  & -2\Gamma & -B_x  \\
0  &  B_x & 0 
\end{array}\right)
\left(\begin{array}{c} s_x \\ s_y \\ s_z \end{array}\right)
\label{Eq:white}
\end{eqnarray}
where $s_z$ is the spin-polarization along the $z$-axis. 
Note that there is no Lamb shift here, so 
the noise does not modify the off-diagonal terms.
One can use the level-repulsion argument to see why the Lamb shift is zero.
If the white-noise were generated by an environment of quantum modes,
all gaps would be equally represented. Thus 
there would be equal weight of modes pushing the system gap
downwards as upwards, leaving the system's gap 
unchanged \cite{footnote:no-Lamb}.
In contrast, for coloured noise
the Lamb shift would be non-zero, and we would
have $B_x$ replaced by $B_x'=B_x+ {\cal O}[\Gamma]$.

It is easy to rewrite Eq.~(\ref{Eq:white}) to show that $s_z$ is given by
the equation of motion of a damped harmonic oscillator; 
$\ddot s_z +\Gamma \dot s_z +B_x^2 s_z=0$.
Thus it has two regimes,
\begin{eqnarray}
\hbox{ Under-damped } (B_x > \Gamma): & & 
s_z(t) = 
  A_+ \e^{-[\Gamma + \rmi(B_x^2-\Gamma^2)^{1/2}]t} 
+ A_- \e^{-[\Gamma - \rmi(B_x^2-\Gamma^2)^{1/2}]t}\qquad
\\
\hbox{ Over-damped } (B_x < \Gamma): & & 
s_z(t) = 
  A_+ \e^{-[\Gamma + (\Gamma^2-B_x^2)^{1/2}]t} 
+ A_- \e^{-[\Gamma - (\Gamma^2-B_x^2)^{1/2}]t} \qquad
\end{eqnarray}
where $A_+$ and $A_-$ are given by $t=0$ boundary conditions.
From this we see that the frequency of coherent oscillations is reduced
in the under-damped regime. By increasing $\Gamma$ one enters the 
over-damped regime. As we take $\Gamma \to \infty$, the second term in 
$s_z$ decays at a rate $\propto \Gamma^{-1}$; thus strong noise stops
transitions of the spin.
Consider the spin as initially  $\left|\up\right\rangle$ along the $z$-axis,
so $s_z(t=0)=1$ and $\dot s_z(t=0)\propto s_y(t=0)=0$,
thus for large $\Gamma$ we have $A_-=1$ and $A_+=0$.
Then $s_z(t) \simeq \exp[-B_x^2t/(2\Gamma)]$, so the transition rate is suppressed by strong noise.
In reviewing Ref.~\cite{Blanchard}, 
Michael Berry \cite{Berry-reviews-Blanchard} 
pointed out that this suppression of transitions in the over-damped case 
was a quantum Zeno effect.  Indeed, in the limit
$\Gamma \ll B_x$, the result coincides with that of
the usual (observation-induced) quantum Zeno effect if we identify 
$\Gamma$ with $2\tau_{\rm obs}^{-1}$.

All the above physics (under-damping and over-damping) is contained
in the Bloch-Redfield equation, which is applicable whenever
the decay time, $1/\Gamma$, is much greater than the environment memory-time
(the memory time is zero for white-noise,
so Bloch-Redfield becomes essentially exact) \cite{whitney07}. 
However often one makes an additional secular or rotating wave-approximation
(e.g. as in Ref.~\cite{wmsg}), which is only justifiable in the 
strongly under-damped limit (where one keeps ${\cal O}[\Gamma]$-terms 
but neglects ${\cal O}[\Gamma^2/B_x]$-terms).
As we see above, once one is beyond this under-damped limit,
level-repulsion is {\it not} the only source of shifts of the 
coherent oscillation frequency.

\vskip 3mm
{\bf A  ``super'' Zeno effect due to the orthogonality catastrophe :}
The orthogonality catastrophe is a sort of extreme Lamb shift,
which can only occur in a system strongly coupled to it environment.
It was called ``adiabatic renormalization'' by 
Leggett {\it et al} \cite{Leggett}, but Anderson introduced the idea earlier
in the context of a fermionic (rather than bosonic) environment \cite{orthog}.
Each mode of the environment tries to adiabatically follow
the system (spin),  which means that all environment
modes must shift each time the spin flips.
Then the transition rate for the spin becomes
$B_x \prod_{n=1}^N \langle n_-|n_+\rangle$, where $\left| n_\pm \right\rangle$ is the
state of the $n$th environment mode shifted one way or the other
(depending if the spin is up or down).  The overlap  $\langle n_-|n_+\rangle$
is slightly less than one, however the product over all 
overlaps will be {\it exponentially small} when the number of environment modes, 
$N\gg 1$.
This argument neglects non-adiabatic effects
(transitions of environment modes).  Leggett {\it et al} \cite{Leggett}
used renormalization group (RG)  and find that they too are exponentially small.
We call this a ``super Zeno effect'' since one 
would have to wait an exponentially long time for the system to undergo a transition.
This timescale is of order $1/(B_x\e^{-\Gamma/\Om})$ where $\Om$ is a typical frequency of the environment (cf. the examples above, where this timescale is of order $1/(B_x^2\tau_{\rm obs})$ 
or $\Gamma/B_x^2$).

However our recent work has called into question the so-called ``irrelevant'' operators in the RG flow.
We suspected that they summed to give a contribution to the transition rate which was not exponentially small, but instead went like an inverse power of the upper cut-off on the environment spectrum, $\Om_{\rm U}$ (here we restrict our comments to super-Ohmic baths) .  
In many cases this cut-off is not so different from other energyscales in the problem,
then the transition rate of the system might be dominated by a
$B_x^2(\Gamma \Om_{\rm U})^{1/2}$-term,
rather than the exponentially suppressed rates found from the adiabatic RG analysis.
To check the contribution of these ``irrelevant'' terms, one would have to keep track of their evolution
under the RG flow, and then sum them.  This is difficult, so instead we have attacked the problem
with a completely different technique.

We take to heart the message of the orthogonality catastrophe
and transforming the original Hamiltonian 
(spin-half coupled to many environment modes) 
to a basis of shifted-environment modes.  
This powerful and elegant approach is standard for polarons \cite{Mahan},
and was first applied to the above problem 
in a number of much neglected papers \cite{Aslangul86,Dekker87,Aslangul88}.
These works showed that the non-interacting blip approximation of 
Ref.~\cite{Leggett} is given by a simple weak-coupling analysis of the 
transformed Hamiltonian.
Elsewhere, we will present a detailed review of this approach 
and discuss its regime of validity (greatly over-estimated 
in Ref.~\cite{Aslangul88}); here we simply note the remarkable
conclusion that the polaron transformation can map a highly non-Markovian environment
onto an almost Markovian one.
Treating the transformed problem with a Bloch-Redfield analysis,
we confirm the existence of  $B_x^2/\Om_{\rm U}$-terms that the RG analysis misses.
They make an important contribution to observables related to the $x$- and $y$-axis spin-polarizations,
as such they are absolutely crucial for analysis of phase information, such as the Berry phase \cite{whit-BP} (note that in earlier versions of \cite{whit-BP}, we used the RG method and so 
missed the terms that are {\it not} exponentially suppressed).

However from our work in progress,
it appears that the $B_x^2(\Gamma \Om_{\rm U})^{1/2}$-terms exactly cancel for the quantity 
considered by Leggett {\it et al}; the $z$-axis spin polarization.
We do not yet know if there is a physical reason behind this cancellation. 
We also do not know if such a cancellation will occur at higher-order in perturbation theory
(the Bloch-Redfield analysis involves a Born approximation). 
If this cancellation occurs at all orders, then the transition rate will decay exponentially with $\Gamma$
for all large $\Gamma$.  However if not, we suspect that for extremely large $\Gamma$, there will be a cross-over to something that goes like some inverse power of  $\Gamma\Om_{\rm U}$.

\section{Using noise to make better Berry phases}

In Ref.~\cite{whit-BP}, we show that a noise-induces
orthogonality catastrophe can cause a 
spin to acquires a  Berry phase when the axis that the noise couples to 
is rotated around a closed loop.
Earlier works \cite{previous-noise-BPs} predicted noise-induced Berry phases
which come from over-damped spin-dynamics,
rather than an orthogonality catastrophe.
We argue that in our case (unlike those earlier works), 
the Berry phase is relatively weakly affected by
decoherence and the leading non-adiabatic phase drops out.
Thus we reach the counter-intuitive conclusion that {\it environmental
noise may make it easier to accurately measure a Berry phase}.

Berry phases occur in many quantum systems 
\cite{Berry84,Anandan-review}, and have potential applications in both 
quantum computation \cite{Jones00Ekert00} and metrology \cite{metrology}.
However in general, Berry phases do not appear ``alone''.
If the parameters of a system's 
Hamiltonian complete a closed loop
in a time $t_{\rm p}$, then the total phase acquired by an eigenstate is
\begin{eqnarray}
\Phi_{\rm total} 
= \Phi_{\rm dyn} + \Phi_{\rm Berry} 
 +\Phi_{\rm NA}^{(1)}
+ \Phi_{\rm NA}^{(2)} + \cdots,
\label{eq:usual-total-phase}
\end{eqnarray} 
where the dynamic phase $\Phi_{\rm dyn}\propto Et_{\rm p}$,
the Berry phase $\Phi_{\rm Berry}\propto (Et_{\rm p})^0$,
and the non-adiabatic (NA) correction $\Phi_{\rm NA}^{(\mu)} 
\propto (Et_{\rm p})^{-\mu}$, with $E$ being a system energy scale 
(such as the gap to excitations).
The Berry phase is hard to isolate 
(and thus hard to use for quantum computation or metrology), because it
is the second term in
this $t_{\rm p}^{-1}$-expansion of $\Phi_{\rm total}$.
To suppress the non-adiabatic terms, one must make $t_{\rm p}$ large.  
This means $\Phi_{\rm dyn} \gg \Phi_{\rm BP}$, so 
one must subtract off $\Phi_{\rm dyn}$
(using a spin-echo trick or degenerate states \cite{Wilczek-Zee}) 
with extreme accuracy. 
For example, to get $\Phi_{\rm BP}$ with an error of 0.1\%,
one must make $Et_{\rm P} \sim 10^3$, which means that 
$\Phi_{\rm dyn} \sim 10^3\,\Phi_{\rm BP}$. 
Then since one wants $\Phi_{\rm BP}$ with an error of 0.1\%, one has to 
subtract off $\Phi_{\rm dyn}$ with an accuracy of 0.0001\%.

In contrast, for the sistuation we consider, the total phase is; 
\begin{eqnarray}
\Phi_{\rm total} 
= \Phi_{\rm Berry} 
+ \Phi_{\rm NA}^{(2)} + \cdots,
\label{eq:usual-total-phase}
\end{eqnarray}  
since there is no dynamic phase and $\Phi_{\rm NA}^{(1)}$ is imaginary
and so no longer contributes to the phase (in some cases it is also exponentially small).
 This can be shown by transforming to a
rotating frame \cite{Berry87}
which follows the slowly varying noise-axis.  Then we arrive at a 
problem similar to that discussed in the previous section of the article.
In this case to get the Berry phase with an error of only 0.1\%,
one simply needs to ensure that $\Phi_{\rm NA}^{(2)}\sim 10^{-3}$,
which requires that $Et_{\rm p} \sim 10^{3/2} \sim 31$.
This requirement is much less strict than that for the convention Berry phase above.

In conclusion, we note that we are currently looking at whether 
such noise can also make non-Abelian geometric phases (Wilczek-Zee phases \cite{Wilczek-Zee}).
Non-Abelian means ``${\cal A}$ then ${\cal B}$'' is different
from ``${\cal B}$ then ${\cal A}$''.  In the above case, 
it appears possible that  
if we perform two different loops (${\cal A}$ and ${\cal B}$) with the noise, 
we get two different (SU(2)) geometric phases depending on which loop 
is followed first.

We thank M.V.~Berry for comments on Ref.~\cite{whit-BP}
and for informing us of Refs.~\cite{Blanchard,Berry-reviews-Blanchard}.



\begin{thebibliography}{1}
\bibitem{footnote:env-increase-P}
The ground state is pure, $P=1$, so environments can enhance
(as well as reduce) the system's purity.
\bibitem{footnote:no-Lamb}
Alternatively one can take a Bloch-Redfield approach,
see for example Eq.~(9) of Ref.~\cite{wmsg}
(dropping the time-dependence and setting $\theta=\pi/2$).
We see immediately that $\Gamma_0$ is real for white-noise 
($S(\Om)=$constant), thus there is no Lamb shift.
\bibitem{footnote:close-to-zero}
$\Phi_{\rm dyn}$ goes linearly in the static B-field. 
In most qubits this can be tuned to close to zero.
\end{thebibliography}
\end{document}